\documentclass[11pt]{article}
\usepackage{epsfig}
\usepackage{enumerate}

\def\be{\begin{equation}}
\def\ee{\end{equation}}
\def\bea{\begin{eqnarray}}
\def\eea{\end{eqnarray}}
\newcommand{\lsim}{\mathrel{\mathop{\kern 0pt \rlap
  {\raise.2ex\hbox{$<$}}}
  \lower.9ex\hbox{\kern-.190em $\sim$}}}
\newcommand{\gsim}{\mathrel{\mathop{\kern 0pt \rlap
  {\raise.2ex\hbox{$>$}}}
  \lower.9ex\hbox{\kern-.190em $\sim$}}}

\newcommand{\AmS}{{\protect\the\textfont2
  A\kern-.1667em\lower.5ex\hbox{M}\kern-.125emS}}
\normalsize
\begin{document}

\baselineskip=0.65cm

\begin{center}
\Large
{\bf No role for muons \\
in the DAMA annual modulation results} \\ 
\vspace{0.5cm}

\rm
\end{center}

\large

\begin{center}

R.\,Bernabei$^{1,2}$,~P.\,Belli$^{2}$,~F.\,Cappella$^{3,4}$,~V.\,Caracciolo$^{5}$,~R.\,Cerulli$^{5}$,
\vspace{1mm}

C.J.\,Dai$^{6}$,~A.\,d'Angelo$^{3,4}$,~A.\,Di Marco$^{1,2}$,~H.L.\,He$^{6}$,~A.\,Incicchitti$^{4}$,
\vspace{1mm}

X.H.\,Ma$^{6}$,~F.\,Montecchia$^{2,7}$,~X.D.\,Sheng$^{6}$,~R.G.\,Wang$^{6}$ and Z.P.\,Ye$^{6,8}$
\vspace{1mm}

\normalsize

\vspace{0.4cm}

$^{1}${\it Dip. di Fisica, Universit\`a di Roma ``Tor Vergata'', I-00133 Rome, Italy}
\vspace{1mm}

$^{2}${\it INFN, sez. Roma ``Tor Vergata'', I-00133 Rome, Italy}
\vspace{1mm}

$^{3}${\it Dip. di Fisica, Universit\`a di Roma ``La Sapienza'', I-00185 Rome, Italy}
\vspace{1mm}

$^{4}${\it INFN, sez. Roma, I-00185 Rome, Italy}
\vspace{1mm}

$^{5}${\it Laboratori Nazionali del Gran Sasso, I.N.F.N., Assergi, Italy}
\vspace{1mm}

$^{6}${\it IHEP, Chinese Academy, P.O. Box 918/3, Beijing 100039, China} 
\vspace{1mm}

$^{7}${\it Laboratorio Sperimentale Policentrico di Ingegneria Medica, Universit\`a
degli Studi di Roma ``Tor Vergata''}
\vspace{1mm}

$^{8}${\it University of Jing Gangshan, Jiangxi, China}
\vspace{1mm}

\end{center}

\normalsize

\begin{abstract}

This paper gathers arguments and reasons why muons surviving the Gran Sasso mountain 
cannot mimic the Dark Matter annual modulation
signature exploited by the DAMA/NaI and DAMA/LIBRA experiments.
A number of these items have already been presented in individual papers.
Further arguments have been addressed here in order to present a comprehensive 
collection and to enable a wider community to correctly approach this point.

\end{abstract}

\section{Introduction}
\label{intro}

The DAMA/NaI and DAMA/LIBRA experiments at the Gran Sasso underground laboratory (LNGS) of the I.N.F.N.
have been and are, respectively, 
investigating the presence of the Dark Matter (DM) particles in the galactic halo by exploiting the model independent DM 
annual modulation signature, 
originally suggested in the middle of '80s in ref. \cite{Freese}.
In fact, as a consequence of the Earth annual revolution around the Sun, 
which is moving in the Galaxy
traveling with respect to the Local Standard of Rest towards
the star Vega near the constellation of Hercules,
the Earth should be exposed to a higher flux of Dark Matter particles around $\sim$ 2 June
(when the Earth orbital velocity is added to the one of the solar system with respect
to the Galaxy) and by a smaller one around $\sim$ 2 December 
(when the two velocities are subtracted).

This DM annual modulation signature is very distinctive since the effect induced by DM
particles must simultaneously satisfy all the following requirements:

\begin{itemize}

\item[I)]  the event rate must contain a component modulated according to a cosine function; 

\item[II)] with period equal to one year; 

\item[III)] with a phase roughly around June 2$^{nd}$ in case of usually adopted halo 
            models (slight variations may occur in case of presence of non thermalized 
            DM components in the halo);

\item[IV)] this modulation must be present only at low energy,
           where DM particles can induce signals; 

\item[V)]  it must be present only in those events where just a single detector, 
           in a multi-detector set-up, actually ``fires'' ({\it single-hit} events), 
           since the probability that DM particles experience multiple 
           interactions is negligible; 

\item[VI)] the modulation amplitude in the region of maximal sensitivity has to be $\lsim$~7\% 
      in case of usually adopted halo distributions, but it may be significantly larger in 
      some particular scenarios.

\end{itemize}

Thus, this signature has a different origin and peculiarities than 
effects correlated with seasons on the Earth.

To mimic such a signature spurious effects or side reactions should be able 
not only to account for the observed 
modulation amplitude but also to simultaneously satisfy all the requirements of the signature; 
thus, no other effect investigated so far in the field of rare processes offers
a so stringent and unambiguous signature.

Let us now briefly describe the DAMA/LIBRA experiment \cite{perflibra}, recalling its model independent 
annual modulation results \cite{modlibra,modlibra2}. The present 
DAMA/LIBRA set-up, installed at the Gran Sasso underground laboratory, 
is made of 25 highly radiopure NaI(Tl) crystal scintillators 
in a 5-rows 5-columns matrix. Each NaI(Tl) detector has 9.70 kg mass and a size 
of ($10.2 \times 10.2 \times 25.4$) cm$^{3}$. 
The scintillation light (decay time $\simeq 240$ ns) of each crystal is collected 
(through two 10 cm long highly radiopure quartz light guides, 
which also act as optical windows being directly coupled to the bare crystal)
by two low-background photomultipliers working in coincidence at single photoelectron threshold.
The software energy threshold in the present data taking is 2 keV electron equivalent (hereafter keV)
and the measured light response is 5.5--7.5 photoelectrons/keV depending on the detector.
In order to reject afterglows, Cherenkov pulses in the
light guides and Bi-Po events, a 500 $\mu$s veto occurs after each event \cite{perflibra}.
The detectors are housed in a low radioactivity sealed 
copper box installed in the center of a low-radioactive Cu/Pb/Cd-foils/polyethylene/paraffin shield;
moreover, about 1 m concrete (made from the Gran Sasso rock material) almost fully surrounds (mostly outside the 
barrack) this passive shield, acting as a further neutron moderator.
In particular, the neutron shield reduces by a factor 
larger than one order of magnitude the thermal neutrons flux \cite{modlibra}.
The copper box is maintained in HP Nitrogen atmosphere in slight overpressure with respect to the 
external environment; it is part of the 3-levels sealing system which prevents environmental air
reaching the detectors.
The experiment takes data up to the MeV scale despite the optimization is made for the lowest energy region.
The linearity and the energy resolution of the detectors at low and high energy have been investigated using
several sources as discussed in ref. \cite{perflibra}; routine calibrations are carried out 
in the same conditions as 
the production runs, by using the glove-box installed in the upper part of the apparatus \cite{perflibra}.

The DAMA/LIBRA data released so far correspond to
six annual cycles for an exposure of 0.87 ton$\times$yr \cite{modlibra,modlibra2}.
Considering these data together with those previously collected by the former DAMA/NaI
over 7 annual cycles (0.29 ton$\times$yr) \cite{RNC,ijmd}, the total exposure collected
over 13 annual cycles is 1.17 ton$\times$yr. 
Several analyses on the model-independent DM annual
modulation signature have been performed (see Refs.~\cite{modlibra,modlibra2} and references therein).
A clear modulation is present in the (2--6) keV {\it single-hit} events and fulfills all the requirements of the 
DM annual modulation signature. In particular, no modulation is observed either above 6 keV or in the
(2--6) keV {\it multiple-hits} events.

The results provide a model
independent evidence of the presence of DM particles in the galactic halo at 8.9 $\sigma$ C.L.
on the basis of the investigated DM signature.
In particular, with the cumulative exposure the modulation
amplitude of the {\it single-hit} events in the (2--6) keV energy
interval, measured in NaI(Tl) target, is $(0.0116 \pm 0.0013)$
cpd/kg/keV; the measured phase is $(146 \pm 7)$ days (corresponding to May 26 $\pm 7$ days) and the
measured period is $(0.999 \pm 0.002)$ yr, values well in agreement
with those expected for the DM particles.

Careful investigations
on absence of any significant systematics or side reaction able to
account for the measured modulation amplitude and to simultaneously satisfy
all the requirements of the signature
have been quantitatively carried out (see e.g. ref.
\cite{perflibra,modlibra,modlibra2,RNC,ijmd,scineghe09,taupnoz,vulca010,canj11,tipp11}, refs therein);
none has been found or suggested by anyone over more than a decade. In particular, the case
of muons has been deeply investigated.

This paper will further demonstrate that 
neither muons nor muon-induced events can significantly contribute to 
the DAMA observed annual modulation signal. In addition, some of the 
already-published arguments \cite{perflibra,modlibra,modlibra2,RNC,ijmd,scineghe09,taupnoz,vulca010,canj11,tipp11} 
are summarized here.

\section{The muon flux at LNGS}   \label{muon}

The muons surviving the coverage of the Gran Sasso laboratory 
either can have direct interactions in the experimental set-up or 
can produce in the surroundings and/or inside the set-up
secondary particles, such as fast neutrons, $\gamma$'s,
electrons, spallation nuclei, hypothetical exotics, etc., possibly depositing energy in the detectors.
Such direct or indirect events are a potential background for low count rate
experiments, as DAMA is. In this paper, the muon induced background in DAMA/LIBRA 
will be investigated and any possible role in the DAMA results will be quantitatively
ruled out.

The surviving muon flux ($\Phi_\mu$) has been measured in the deep
underground Gran Sasso Laboratory (3600 m w.e. depth) 
by various experiments with very large exposures \cite{Mac97,LVD,borexino,borexino2};
its value is $\Phi_\mu \simeq 20 $ muons m$^{-2}$d$^{-1}$ \cite{Mac97},
that is about a factor $10^6$ lower than the value measured at sea level.
The measured average single muon energy at the Gran Sasso laboratory is
$270 \pm 3 (stat) \pm 18(syst)$ GeV \cite{Mac03}; this value agrees with 
the predicted values using different parametrizations \cite{hime}.
A $\simeq$ 2\% yearly variation of the muon flux was firstly measured years ago by
MACRO; when fitting the data of the period January 1991 -- December 1994 all together,
a phase around middle of July was obtained \cite{Mac97}.
It is worth noting that 
the flux variation of the muons is attributed to the
variation of the temperature in the outer atmosphere, and its phase
changes each year depending on the weather condition.
Recently, other measurements have been 
reported by LVD, quoting a lower amplitude (about $1.5\%$) and
a phase, when considering the data of the period January 2001 -- December 2008 all together,
equal to (5 July $\pm$ 15 days) \cite{LVD}. Finally, Borexino,
has quoted a phase of (7 July $\pm$ 6 days), still
considering the data taken in the period May 2007 -- May 2010 all together \cite{borexino}.
More recently, the Borexino collaboration presented a modified phase evaluation 
(29 June $\pm$ 6 days)\footnote{It is worth noting that in ref. \cite{borexino2}
28 June (179.0 days) is instead quoted as measured phase; actually, 
in our convention -- coherent throughout the paper -- 179.0 days correspond to 00:00 of 
of 29 June (as, for example, $t=0.0$ days is 00:00 of 1st of January and $t=1.0$ days is 00:00 of 2nd of January).}, with a
still lower modulation amplitude: about $1.3\%$ \cite{borexino2}, by adding the data collected in a further year;
the appreciable difference in the fitted values further demonstrates the large variability 
of the muon flux feature year by year.

\section{Why muons cannot play any role}

The measured muon variation at LNGS has no impact on the DAMA annual modulation results,
recalled in Sect. \ref{intro}. In the following sections we summarize the key items.
It is worth noting
the arguments reported in ref. \cite{chang11}, where no evidence for a correlation
between cosmic rays and DAMA result has been found and it is shown that the two phenomena
differ in their power spectrum, phase, and amplitude.

\subsection{Intrinsic inability of muons to mimic the DM annual modulation signature}
\label{lab1}

Let us firstly recall
\cite{perflibra,modlibra,modlibra2,RNC,ijmd,scineghe09,taupnoz,vulca010,canj11,tipp11} that 
a muon flux variation cannot mimic the Dark Matter annual 
modulation signature in DAMA/LIBRA (and even less in the smaller
DAMA/NaI) set-up, not only  because it may give rise just 
to quantitatively negligible effects (see later for details), 
but also because it is unable to mimic the DM signature. 
In fact, it would fail some requirements of the signature; namely e.g.: 

   \begin{itemize}
   \item[i)]  it would induce variation in the whole energy spectrum.

   \item[ii)]   it would induce variation in the
        {\it multiple-hits} events (events in which more than one detector ``fires''), 

   \item[iii)]  it would induce variation with a phase and amplitude
        distinctively different from the DAMA measured one (see later).
   \end{itemize}

\subsection{Inconsistency between the phase of muons and of the muon-induced effect and the DAMA phase}   \label{phase}

\begin{figure}[!ht]
\vspace{.4cm}
\centering
\includegraphics[width=0.75\textwidth] {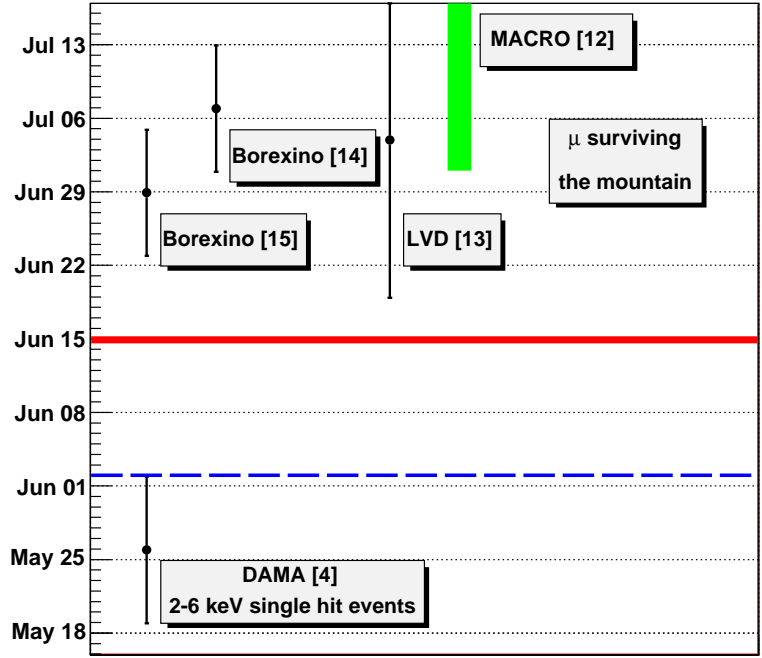}
\vspace{-.4cm}
\caption{The phase of the DAMA annual modulation signal \cite{modlibra2} 
and the muon phases quoted by Borexino in two analyses (May 2007 -- May 2010 \cite{borexino},
and May 2007 -- May 2011 \cite{borexino2}),
by LVD (January 2001 -- December 2008 \cite{LVD}), and by MACRO (January 1991 -- December 1994 \cite{Mac97}).
The muon phases quoted by those three experiments have 
to be regarded as mean values among the muon phases in all the considered years since the muon phase 
depends on the temperature of the outer atmosphere and, thus, it changes each year.
The phase of the DAMA observed effect has instead a stable value in the different years 
\cite{modlibra,modlibra2}. The 
horizontal dashed line corresponds to 2nd June (date around which the phase of the DM annual modulation is 
expected). The middle of June is also marked as an example; in fact, the maximum temperature of the $T_{eff}$
at the LNGS location (see text) cannot be as early as the 
middle of June (and for several years), date which is still 3 $\sigma$ far away from the phase of the DAMA 
observed effect.}
\label{fg:phase}
\end{figure}

The phase of muons surviving the Gran Sasso coverage, measured deep underground at LNGS,
and the phase of the (2--6) keV {\it single-hit} events measured by DAMA
are distinctively different (see Fig. \ref{fg:phase}).
In particular, the values quoted by MACRO, LVD and Borexino experiments for the muon phase
have to be regarded as mean values of the muon phases among the analyzed years
and the associated errors are not simply due to statistical fluctuation of the 
data, but rather to the variations of the muon phase in the different years.
The phase of the DAMA observed effect has instead a stable value in the different 
years \cite{modlibra,modlibra2} and is 5.7 (5.9, 4.7) $\sigma$ from the LVD (Borexino, first and recently  
modified evaluations, respectively) ``mean'' phase of muons (7.1 $\sigma$ from the 
MACRO one).

This simple approach does not consider that the experimental errors in the muon flux 
are not completely Gaussian; however, it gives the right order of magnitude of the confidence level
for the incompatibility between the DAMA phase and the phase of muons and of the muon-induced effects.
Analyses carried out by different authors confirm these outcomes; for example, a  
disagreement in the correlation analysis between the LVD data on muon flux and the
DAMA residuals with a confidence level greater than 99.9\% is reported in ref. \cite{chang11}.

It is also worth noting that the expected phase for DM is significantly different than the expected phase 
of muon flux at Gran Sasso: in fact, while the first one is always about 152.5 day of the year, 
the second one is related to the variations of the atmospheric
temperature above the site location, $T_{eff}$.
In particular, the atmosphere is generally considered as
an isothermal body with an effective temperature $T_{eff}$; the behaviour of $T_{eff}$
at the LNGS location as function of time has been determined e.g. in ref. \cite{borexino2}.
As first order approximation $T_{eff}$ was fitted with a cosinusoidal behaviour and the phase
turned out to be (24 June $\pm$ 0.4 days) \cite{borexino2};
this is later than e.g. the middle of June, date     
which is still $3 \sigma$ far away from the DAMA measured phase (see Fig. \ref{fg:phase}).
In addition, fitting
the temperature values at L'Aquila in the years 1990-2011 \cite{tempaq} 
with a cosinusoidal function, a period of $(365.1 \pm 0.1)$ days and a phase
of (25 July $\pm$ 0.6 days) are obtained. 

Thus, in conclusion, the phase of the DAMA annual    
modulation signal \cite{modlibra2} is 
significantly different than the phases of the surface temperature and of the
$T_{eff}$, on which the muon flux is dependent, and than the phases of the muon flux
measured by MACRO, LVD and Borexino experiments.

The above argument also holds for every kind of cosmogenic product
(even hypothetical exotics) due to muons. 

\begin{figure}[!ht]
\centering
\includegraphics[width=0.68\textwidth] {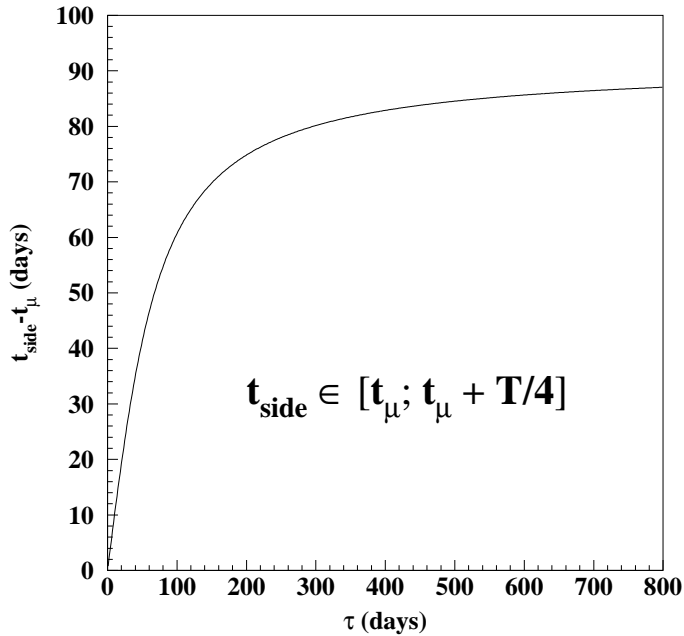}
\vspace{-.4cm}
\caption{Time difference, ($t_{side}$-$t_{\mu}$), as function of $\tau$. 
The $t_{side}$ parameter is the expected phase in case of any contribution 
to the modulation amplitude arising from the decay or the de-excitation (or whatever else) 
of hypothetical cosmogenic product (even exotic) produced by muons
and having mean-life time $\tau$. The muon phase of each considered year
is indicated as $t_{\mu}$. Any exotic contribution related to muons
would have a phase larger than the muon phase and, thus, even more distant
from the expected DM phase and from the DAMA measured phase.
$T$ is the 1 year period; see text.}
\label{fg:tau}
\end{figure}

In particular, when the decays or the 
de-excitations of any
hypothetical cosmogenic product have mean-life time $\tau$,
the expected phase, $t_{side}$,
would be (much) larger than the muon phase (of each considered year) 
$t_{\mu}$, as shown in Fig. \ref{fg:tau}, 
and even more different from the one measured by the DAMA experiments 
and expected from the DM annual modulation signature ($\simeq$ June 2nd).
In fact, the number of the cosmogenic products, $N(t)$, satisfies the following
equation:
\begin{center}
$dN = -N(t)\frac{dt}{\tau} + \left[ a + b \; cos \omega (t-t_\mu) \right] dt$
\end{center}
given by the sum of two contributions, the former due to the decay of the species and the latter
due to their production, showing the typical pattern of muon flux with $b/a \simeq 0.015$, and
period $T=2\pi/\omega=1$ year; $a$ is the mean production rate. 
Solving this differential equation, one has:
\begin{center}
$N(t) = A e^{-t/\tau} + a \tau + \frac{b}{\sqrt{(1/\tau)^2 + \omega^2}} \; cos \omega (t-t_{side}) $
\end{center}
where $A$ is an integration constant, and $t_{side} = t_\mu + \frac{arctg(\omega\tau)}{\omega}$
(see Fig. \ref{fg:tau}).
In condition of secular equilibrium (obtained for time scale greater than $\tau$), 
the first term vanishes and the third term shows an annual modulation pattern 
with phase $t_{side}$. The relative modulation amplitude of the effect is: 
$\frac{b/a}{\sqrt{1 + (\omega\tau)^2}}$.

Two extreme cases can be considered: if $\tau \ll T/2\pi$, one gets $t_{side} \simeq t_\mu + \tau$;
else if $\tau \gg T/2\pi$, one gets $t_{side} \simeq t_\mu + T/4$ $(\simeq t_\mu + 90$ days) and the relative modulation 
amplitude of the effect is $\ll 1.5\%$.

In conclusion, the phase of muons and of whatever (even hypothetical) muon-induced effect is inconsistent
with the phase of the DAMA annual modulation effect.

\subsection{No role for the muons interacting in the detectors directly}

In addition to the previous arguments, the direct interaction of muons crossing the 
DAMA set-ups cannot give rise to any 
appreciable variation of the measured rate. In fact, the 
exposed NaI(Tl) surface of 
DAMA/LIBRA is about 0.13 m$^2$ (and smaller in the former DAMA/NaI); thus
the muon flux in the $\simeq$ 250 kg DAMA/LIBRA set-up is about 2.5 muons/day.
In addition, the impinging muons give mainly {\it multiple-hits} events and over the 
whole energy spectrum.

\begin{figure}[!ht]
\centering
\includegraphics[height=7.5cm] {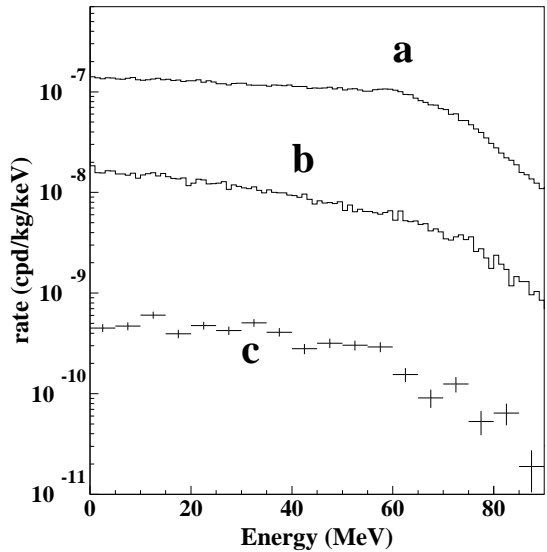}
\vspace{-.3cm}
\caption{{\it Single-hit} event rate as a function of the energy, expected 
for different sets of DAMA/LIBRA detectors from the 
direct interaction of muons crossing the DAMA set-ups, taking into account the muon 
intensity distribution, the Gran Sasso rock overburden map,
and the geometry of the set-up. 
Case a): average contribution of the 5 upper and 5 lower detectors
in the $5\times5$ matrix. Case b):
average contribution of the remaining 15 detectors. Case c) average contribution of
the 9 inner detectors \cite{fond40}.
}
\label{fig_mu}
\end{figure}

The order of magnitude of such a contribution has been estimated by a 
Montecarlo calculation \cite{fond40,papep} which takes into account
the measured muon intensity distribution as function of the incident direction and 
energy, the Gran Sasso overburden map \cite{ambr95},
and the geometry of the set-up \cite{perflibra}.
The direct interaction of muons is simulated according to the impinging direction of muons on 
the detector and to the energy loss per specific path in the detector. 
Three topologies of the detectors' locations in the $5\times5$ DAMA/LIBRA detectors' matrix have been 
considered. The results are shown in Fig.~\ref{fig_mu}.
One can easily infer that the contributions from muons interacting in the detectors 
directly -- for {\it single-hit} 
events in the (2--6) keV energy region -- to the DAMA total counting rate (that is around 1 cpd/kg/keV) 
and to the observed annual modulation amplitude (that is around $10^{-2}$ cpd/kg/keV) 
are negligible (many orders of magnitude lower).
In addition, as mentioned above, this contribution would also fail some requirements 
of the DM annual modulation signature such as III, IV and V.

\subsection{No role for fast neutrons produced by muon interaction}

The surviving muons and the muon-induced cascades or showers can
be sources of neutrons in the underground laboratory.
Such neutrons produced by cosmic rays are substantially
harder (extending up to several hundreds MeV energies \cite{hime}) 
than those from environmental radioactivity;
their typical flux is about $10^{-9}$ neutrons/cm$^2$/s \cite{kud08}, 
that is three orders of magnitude 
smaller than the neutron flux produced by radioactivity.

In particular, the fast neutron rate produced by muons interaction is given by:
      \begin{center}
      $R_n = \Phi_\mu \cdot Y \cdot M_{eff}$,
      \end{center}
where $M_{eff}$ is the effective mass where muon interactions can give rise to 
events detected in the DAMA set-up and $Y$ is the integral neutron yield, 
which is normally quoted in neutrons per muon per g/cm$^2$ of the crossed 
target material. 

The integral neutron yield critically depends on the chemical composition 
and on the density of the medium through which the muons interact. 
The dependence on atomic weight is well described by a power law \cite{hime}:
$Y = 4.54 \times 10^{-5} \; A^{0.81}$ neutrons per muon per g/cm$^2$;
alternatively, it can also be expressed as \cite{hime}:
$Y = 1.27 \times 10^{-4} \; (Z^2/A)^{0.92}$ neutrons per muon per g/cm$^2$.
Thus, the integral yield of neutrons produced 
by muons deep underground at LNGS is 
$Y \simeq (1 - 7) \times 10^{-4}$ neutrons per muon per g/cm$^2$ \cite{Agl,hime}
for relatively light nuclei and 
$Y \simeq 4.5 \times 10^{-3}$ neutrons per muon per g/cm$^2$ \cite{hime}
for lead. 

Consequently, the modulation amplitude of the
{\it single-hit} events in the lowest energy region induced in DAMA/LIBRA by a muon 
flux modulation can be estimated according to:
      \begin{center}
          $S_m^{(\mu)} = R_n \cdot g \cdot \epsilon \cdot f_{\Delta E} \cdot
          f_{single} \cdot 1.5\% / (M_{set-up} \cdot \Delta E)$,
      \end{center}
where $g$ is a geometrical factor, $\epsilon$ is the detection efficiency for neutrons, 
$f_{\Delta E}$ is the acceptance of the considered energy window 
(E $\ge$ 2 keV), $f_{single}$ is the {\it single-hit} efficiency and 1.5\% is the 
muon modulation amplitude.
Since: 
      \begin{center}
       $M_{set-up} \simeq$ 250 kg, \\
       $\Delta E \simeq$  4 keV,
      \end{center}
assuming the very cautious values:
      \begin{center}
      $g \simeq \epsilon \simeq f_{\Delta E} \simeq f_{single} \simeq 0.5$,
      \end{center}
and taking for $M_{eff}$ the total mass of the heavy shield, 15 ton, one obtains:
      \begin{center}
      $S_m^{(\mu)} < (0.3 - 2.4) \times 10^{-5}$ cpd/kg/keV 
      \end{center}
that is, $S_m^{(\mu)} \ll 0.5\%$ of the observed {\it single-hit} events modulation amplitude.
Even assuming in the calculation the yield $Y$ for lead, the upper limit of $S_m^{(\mu)}$ 
($< 1.5 \times 10^{-4}$ cpd/kg/keV) remains still lower 
than 1\% of the observed {\it single-hit} events modulation amplitude.

We stress that -- in addition -- the latter value has been overestimated by orders 
of magnitude both because of the extremely cautious values assumed in the
calculation and because of the omission of the effect of the neutron shield of the 
set-up.

In conclusion, any appreciable contribution from fast neutrons produced by the muon interactions
can be quantitatively excluded.
In addition, it also would fail some of the requirements of the DM annual modulation signature 
such as III, IV and V.

For completeness, in the next two sub-sections we will address the case of environmental neutrons of whatever 
origin (and, thus, also including those induced by muons). In the first sub-section  
the outcomes in Refs. \cite{modlibra,modlibra2,RNC,ijmd,syst} 
will be recalled without entering in details, while 
in the second one the case of the neutron capture on Iodine will be analysed in depth.

\subsection{... and no role for environmental neutrons}   \label{env}

Environmental neutrons cannot give any significant contribution to the annual modulation measured by the DAMA 
experiments \cite{modlibra,modlibra2,RNC,ijmd,syst}.
In fact, the thermal neutron flux surviving the multicomponent DAMA/LIBRA shield has 
been determined by studying the possible presence of $^{24}$Na from neutron activation of $^{23}$Na in NaI(Tl).
In particular, $^{24}$Na presence has been investigated by looking for triple coincidences induced
by a $\beta$ in one detector and by the two $\gamma$'s in two adjacent
ones. An upper limit on the thermal neutron flux surviving
the multicomponent DAMA/LIBRA shield has been
derived as: $< 1.2 \times 10^{-7}$ n cm$^{-2}$ s$^{-1}$ (90\% C.L.) \cite{modlibra}, that is at least one order of magnitude 
lower than the value of the environmental neutrons measured at LNGS. 
The corresponding capture rate is: $< 0.022$ captures/day/kg. 
Even assuming cautiously a 10\% modulation (of whatever origin\footnote{For 
the sake of correctness, it is worth noting that
a variation of the neutron flux in the underground Gran Sasso laboratory
has never been suitably proved. 
In particular, besides few speculations, there is just an unpublished 2003 short internal report of 
the ICARUS collaboration, TM03-01, that seemingly reports a 5\% environmental neutron variation 
in hall C by exploiting the pulse shape discrimination (PSD) in commercial BC501A liquid scintillator.
However, the stability of the data 
taking and of the applied PSD procedures over the whole data taking period and also the nature of 
the discriminated events
are not fully demonstrated. 
Anyhow, even assuming the existence of   
a similar neutron variation, it cannot quantitatively contribute to the DAMA observed modulation 
amplitude \cite{modlibra,RNC,ijmd} as well as satisfy all the peculiarities of the DM annual modulation
signature.}) of the thermal 
neutrons flux, and with the same phase and period as for the DM case, the corresponding modulation amplitude in the 
lowest energy region would be \cite{modlibra,RNC}: $<0.01 \%$ of the DAMA observed modulation amplitude. 
Similar outcomes have also been achieved for the case of fast neutrons; the fast neutrons have been measured in 
the DAMA/LIBRA set-up by means of the inelastic reaction $^{23}$Na$(n,n')^{23}$Na$^*$ (2076 keV)
which produces two $\gamma$'s in coincidence (1636 keV and 440 keV). An upper
limit -- limited by the sensitivity of the method -- has been found:
$< 2.2 \times 10^{-7}$ n cm$^{-2}$ s$^{-1}$ (90\% C.L.) \cite{modlibra},
well compatible with the value measured at LNGS;
a reduction at least an order of magnitude is expected due to the neutron shield of the set-up.
Even when cautiously assuming a 10\% modulation (of whatever origin) of the fast neutrons flux, and with the same phase 
and period as for the DM case, the corresponding modulation amplitude 
is $<0.5 \%$ of the DAMA observed modulation amplitude \cite{modlibra,RNC}. 

Moreover, in no case the neutrons can mimic the DM annual modulation signature since some of the peculiar 
requirements of the signature would fail, such as III, IV and V.

\subsection{No role for $^{128}$I decay}

Ref. \cite{ralston} has claimed that environmental neutrons (mainly thermal and/or epithermal),
occasionally produced by high energy muon interactions, once captured by Iodine might contribute to
the modulation observed by DAMA through the decay of activated $^{128}$I (that produces -- among others -- low 
energy X-rays/Auger electrons). Such an hypothesis is already excluded by 
several arguments given above (as e.g. those in 
Sect. \ref{lab1} and \ref{env}), moreover it has already been rejected in ref.~\cite{vulca010,canj11}; anyhow,
in the following we will focus just on its main argument avoiding to comment on several other wrong 
statements present in ref. \cite{ralston}.

\begin{figure}[!ht]
\centering
\includegraphics[width=0.85\textwidth] {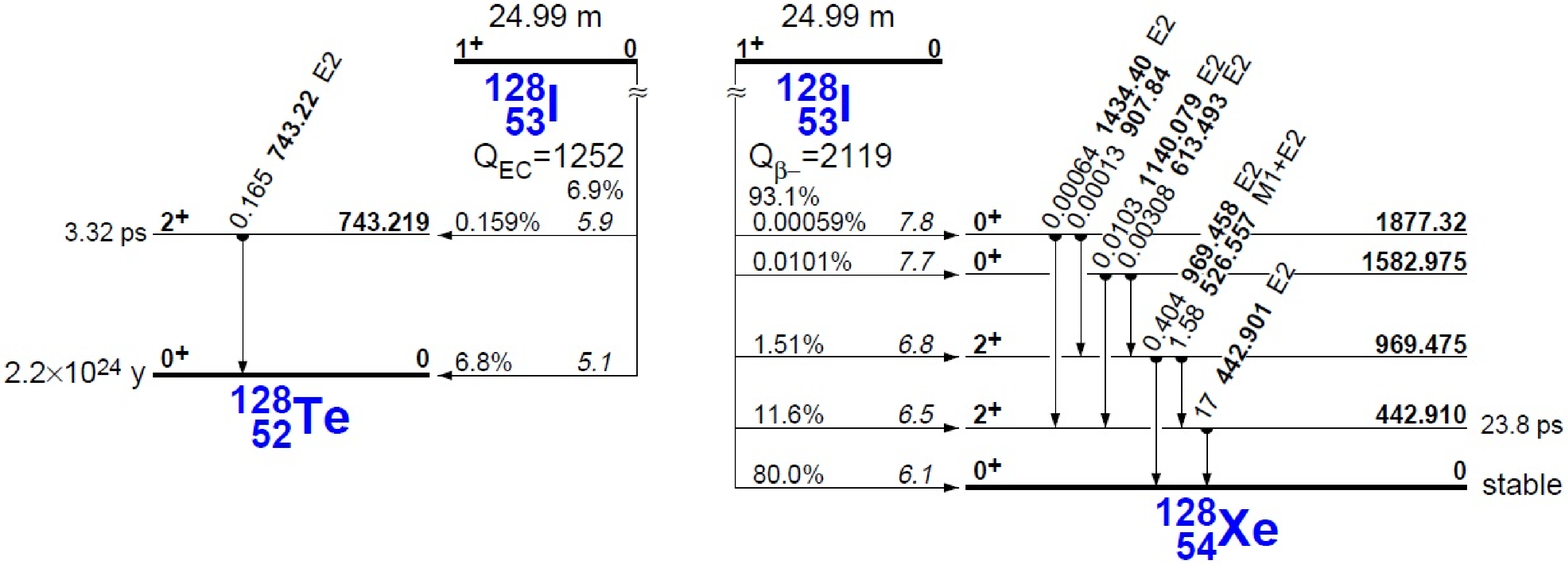}
\vspace{-.2cm}
\caption{The decay schema of $^{128}$I \cite{iod128}.}
\label{fg:128ia}
\end{figure}

The $^{128}$I decay schema is reported in Fig. \ref{fg:128ia}. 
When $^{128}$I decays via the EC channel (6.9\%), it produces low energy 
X-rays and Auger electrons, totally contained inside the NaI(Tl) detectors;
thus, the detectors would measure the total energy release of all the X-rays and Auger electrons, that is 
the atomic binding energy either of the $K$-shell (32 keV) or of the $L$-shells (4.3 to 5 keV) of the 
$^{128}$Te. The probability that so low-energy gamma's and electrons would escape a detector is very small.
In ref. \cite{ralston} it is claimed that such low-energy gamma's and electrons from the $L$ shells may 
contribute to the DAMA observed annual modulation signal; but:

\begin{enumerate}

\item considering the branching ratios of the EC processes in the $^{128}$I decay, the $K$-shell contribution 
(around 30 keV) must be about 8 times larger than that of $L$-shell; while no modulation has been observed 
by DAMA above 6 keV (see \cite{modlibra,modlibra2} and references therein) and, in particular, around 30 keV;

\item the $^{128}$I also decays by $\beta^-$ with much larger branching ratio (93.1\%) than EC (6.9\%) and with 
a $\beta^-$ end-point energy at 2 MeV. Again, no modulation has instead been observed in DAMA experiments at 
energies above 6 keV \cite{modlibra,modlibra2};

\item the $L$-shell contribution would be a gaussian centered around 4.5 keV; this shape is excluded by the 
behaviour of the measured modulation amplitude, $S_m$, as a function of energy (see  Fig. \ref{128I}--{\it Bottom}).
The efficiencies to detect an event per one $^{128}$I decay are: $2 \times 10^{-3}$, $6 \times 10^{-3}$, 
and $2 \times 10^{-3}$ in (2--4) keV, (4--6) keV and (6--8) keV respectively, as calculated by the Montecarlo 
code. Thus, the contribution of $^{128}$I in the (2--4) keV would be similar to the one in the (6--8) 
keV, while the data exclude that.

\end{enumerate}

\begin{figure}[!ht]
\centering
\vspace{-.8cm}
\includegraphics[width=0.6\textwidth] {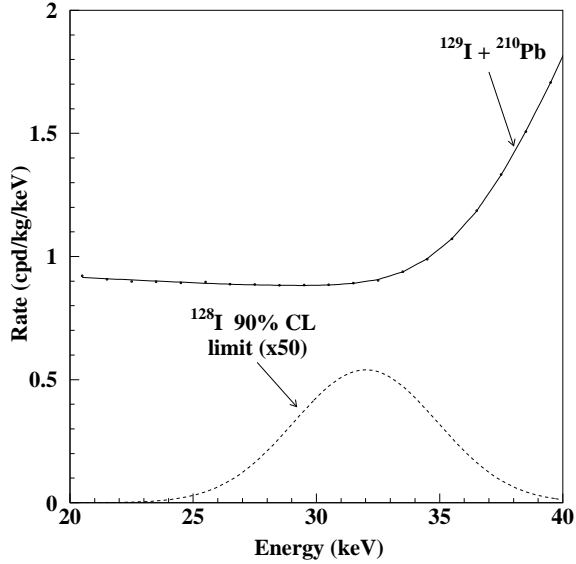}
\vspace{-.4cm}
\caption{Energy distribution of the events measured by DAMA/LIBRA
in the region of interest for the $K$-shell EC decay of $^{128}$I;
the exposure here is 0.53 ton $\times$ yr. The solid line represents the result
of the fit described in ref. \cite{papep}, including the contributions
of $^{129}$I and $^{210}$Pb to the background.
The Gaussian (dashed) line is 50 times the limit of the $^{128}$I contribution,
0.074 cpd/kg, excluded at 90\% C.L.}
\label{fg:128ib}
\end{figure}

Moreover, the data collected by DAMA/LIBRA allow the determination of the possible presence of $^{128}$I in the 
detectors. In fact, neutrons would generate $^{128}$I homogeneously distributed in the NaI(Tl) detectors; therefore 
studying the characteristic radiation of the $^{128}$I decay and comparing it with the experimental data, one can obtain
the possible $^{128}$I concentration. The most sensitive way to perform such a measurement is to study the possible 
presence of the 32 keV peak ($K$-shell contribution) in the region around 30 keV. This was already done and published 
by DAMA -- for other purposes -- in ref. \cite{papep} before ref. \cite{ralston} appeared. 
As it can be observed in Fig. \ref{fg:128ib}, 
there is no evidence of such a peak in the DAMA/LIBRA data; hence an upper limit on the area of a peak around 32 keV 
can be derived to be: 0.074 cpd/kg (90\% C.L.) \cite{papep}. Considering the branching ratio for $K$-shell EC,
the efficiency to detect events in the energy interval around 30 keV for one $^{128}$I decay 
has been evaluated by the Montecarlo code to be 5.8\%.
Hence, one can obtain a limit on possible activity of $^{128}$I ($a_{128}$): 
      \begin{center}
      $a_{128} < 15 \; \mu$Bq/kg (90\% C.L.). 
      \end{center}
\begin{figure}[!ht]
\begin{center}
\includegraphics[width=8.8cm] {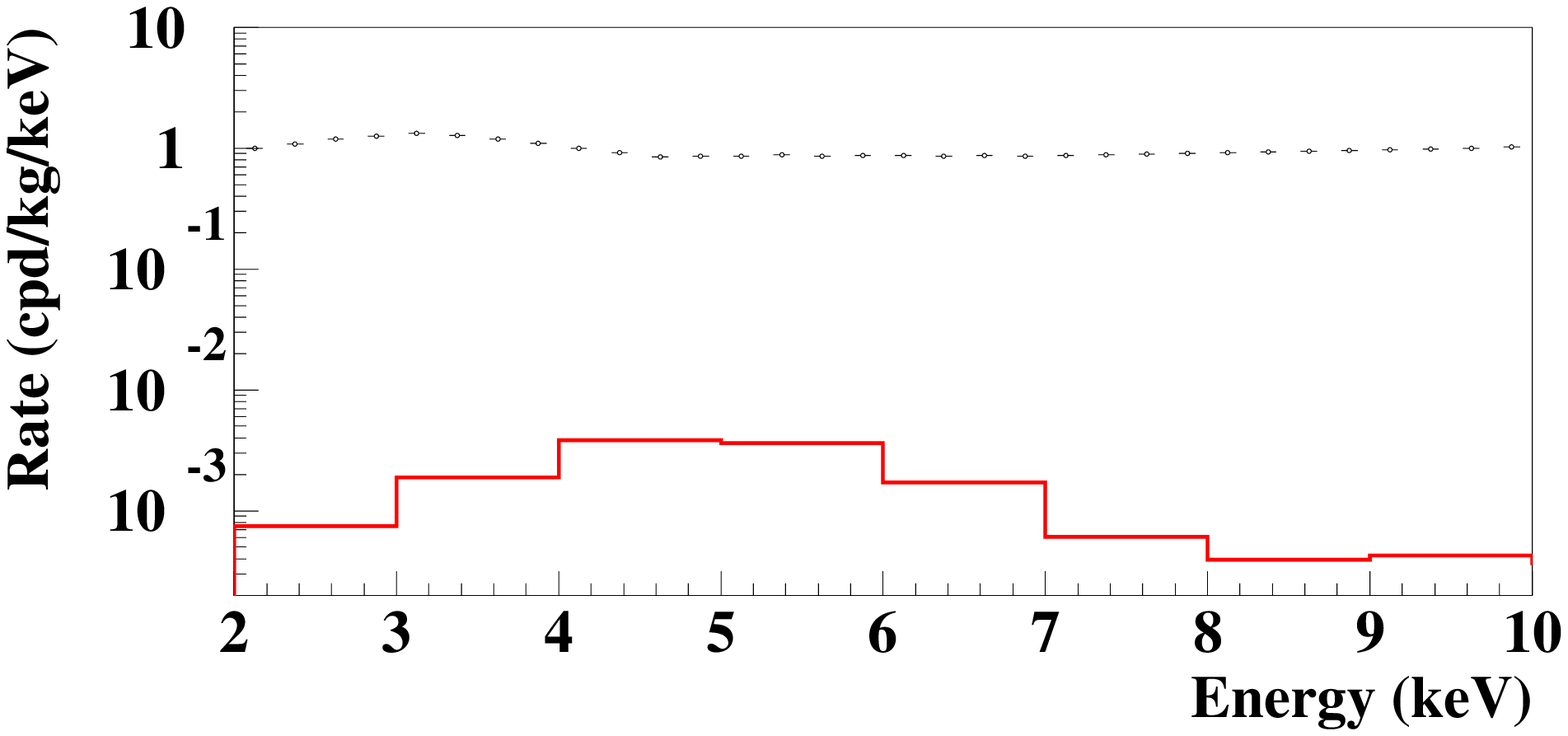}
\includegraphics[width=8.8cm] {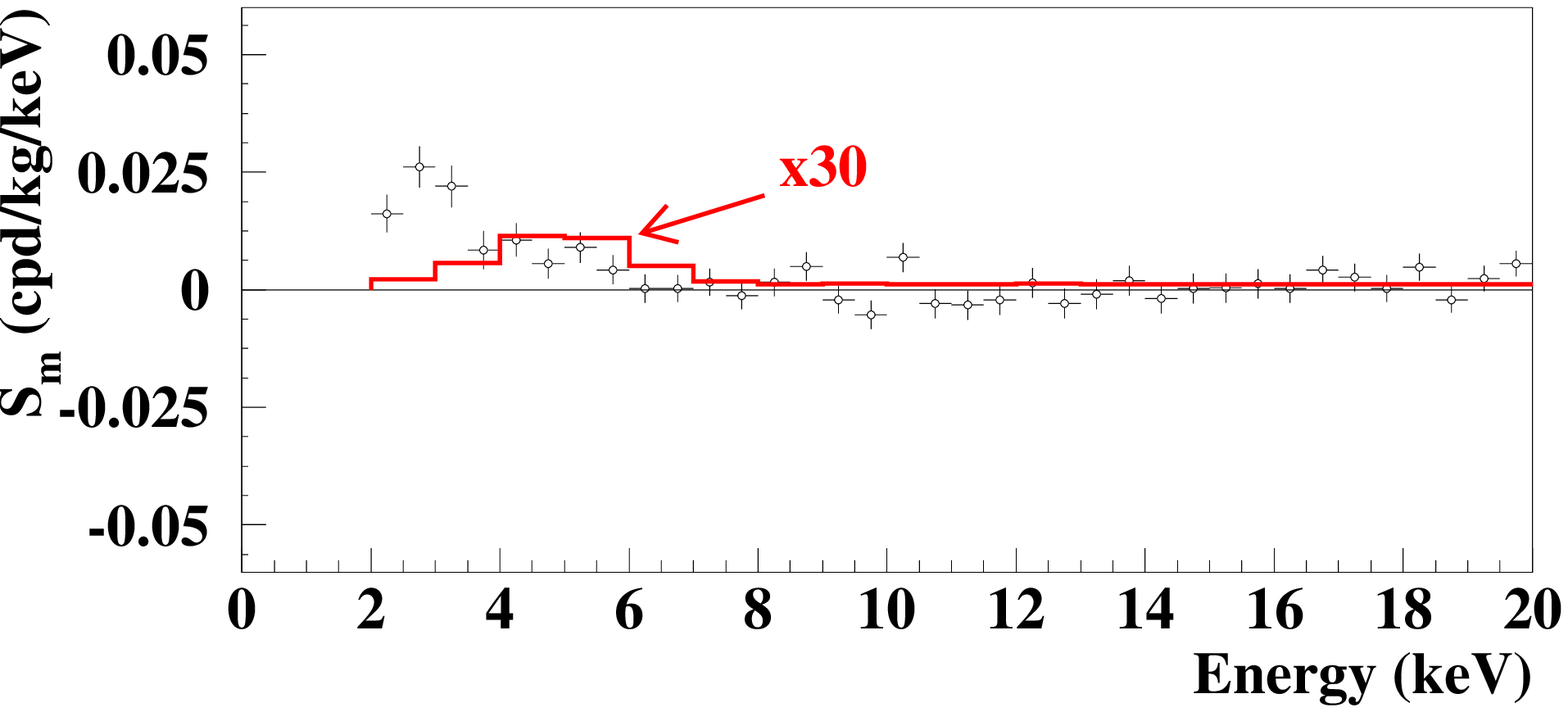}
\end{center}
\vspace{-0.8cm}
\caption{{\it Top -} Data points: cumulative low-energy distribution of the {\it single-hit} scintillation events
measured by DAMA/LIBRA \cite{perflibra} above the 2 keV software energy threshold of the experiment. Histogram (color 
online): maximum expected counting rate from $^{128}$I decays corresponding to the measured upper limit on $^{128}$I 
activity in the NaI(Tl) detectors: $<$15 $\mu$Bq/kg (90\% C.L.); see the data in ref. \cite{papep} and the text.
{\it Bottom -} Data points: the DAMA measured modulation amplitude as a function of the energy. Histogram (color 
online): maximum expected modulation amplitude multiplied by a factor 30 as a function of the energy from $^{128}$I 
decays corresponding to the measured upper limit on $^{128}$I activity given above when assuming an hypothetical 10\% 
neutron flux modulation, and with the same phase and period as a DM signal.}
\label{128I}
\end{figure}
This upper limit allows us to derive the maximum counting rate which may be expected from $^{128}$I in the keV region;
it is reported in Fig. \ref{128I}--{\it Top} together with the cumulative low-energy distribution of the {\it single-hit} 
scintillation events measured by DAMA/LIBRA \cite{perflibra}. It can be noted that any hypothetical contribution from 
$^{128}$I would be negligible. Moreover, even assuming the hypothetical case of a 10\% environmental neutron flux 
modulation, and with the 
same phase and period as the DM signal, the contribution to the DAMA measured (2--6) keV {\it single-hit} modulation 
amplitude would be $<3\times10^{-4}$ cpd/kg/keV at low energy, as reported in Fig. \ref{128I}--{\it Bottom}, that is 
$<2\%$ of the DAMA observed modulation amplitudes. 
In conclusion, any single argument given in this section excludes a role played by $^{128}$I.

\vspace{0.3cm}
\subsection{No role for hypothetical phosphorescence induced by muons}     \label{phos}

In ref. \cite{nygren} it is argued that delayed phosphorescent pulses induced 
by the muon interaction in the NaI(Tl) crystals might contribute to the (2--6) keV 
{\it single-hit} events.
Many wrong statements are put forward in that reference.
We have already critically addressed ref. \cite{nygren} in our ref. \cite{tipp11}.
We will just focus on the main argument.

\vspace{0.3cm}

Since the $\mu$ flux in DAMA/LIBRA is about 2.5 $\mu$/day (see Sect. \ref{muon}), the 
total $\mu$ modulation amplitude in DAMA/LIBRA is about: 
$0.015 \times 2.5$ $\mu$/day $\simeq 0.0375$ $\mu$/day (1.5\% muon modulation 
has been adopted, see Sect. \ref{muon}).
The {\it single-hit} modulation amplitude measured in DAMA/LIBRA 
in the (2--6) keV energy interval is instead \cite{modlibra,modlibra2}:
\begin{center}
$S_{m}(2-6 \; \textrm{keV})
\times \Delta E \times M_{set-up} \sim 10^{-2} \; \textrm{cpd/kg/keV} \times 
4 \; \textrm{keV} \times 250 \; \textrm{kg} \sim 10 \;\textrm{cpd}$.
\end{center}
Thus, the number of muons is too low to allow a similar effect to contribute 
to the DAMA observed amplitude; in fact, to give rise to the DAMA 
measured modulation amplitude each $\mu$ should give rise to about 
270 ($\sim$ 10 counts/day / (0.0375 $\mu$/day), see above) {\it single-hit} correlated events in the 
(2--6) keV energy range in a relatively short period. But:

\begin{itemize}

\vspace{0.2cm}
\item[i)] such a hypothesis would imply dramatic consequences for every NaI(Tl) 
detector at sea level (where the $\mu$ flux is $10^{6}$ times larger than deep underground at 
LNGS), precluding its use in nuclear and particle physics;

\vspace{0.2cm}
\item[ii)] phosphorescence pulses (as afterglows) are single and spare photoelectrons with
very short time decay ($\sim$10 ns); they appear as ``isolated'' uncorrelated spikes. On the other
side, scintillation 
events are the sum of correlated photoelectrons following the typical time distribution 
with mean time equal to the scintillation decay time ($\sim$240 ns).
Pulses with short time structure are already identified and rejected in the noise rejection procedure
described in detail in \cite{perflibra} (the information on the pulse profile is recorded).
Thus, in addition, phosphorescence pulses are not present in the DAMA annual modulation data;

\item[iii)] because of the poissonian fluctuation on the number of muons, the standard deviation of 
the (2--6) keV {\it single-hit} modulation amplitude due to a similar effect would be 13 times 
(see Appendix)
larger than that measured by DAMA, and therefore no statistically-significant effect, produced by 
any correlated events, could be singled out.
Even just this argument (that will be further illustrated in the following) is enough to discard 
the hypothesis of ref. \cite{nygren} (similar considerations are also reported in ref. \cite{blum});

\item[iv)] the muon phase is inconsistent with the phase measured by DAMA (see Sect. \ref{phase}).

\end{itemize}

\vspace{0.2cm}

Thus, the argument regarding a possible contribution from delayed
phosphorescent pulses induced by muons can be safely rejected.

\vspace{0.4cm}
\subsection{Absence of long term modulation}

In ref. \cite{blum} it is argued that high-energy muons measured by LVD might show a long term 
modulation with a period of about 6 years, suggesting that a similar 
long term modulation might also be present in the DAMA data.
We avoid to comment here on the other arguments reported in ref. \cite{blum} 
that are actually already addressed elsewhere in the present paper, and already discard such a suggestion.
However, for completeness such a long term modulation has also been looked for in the DAMA data.

\begin{figure}[!ht]
\begin{center}
\includegraphics[width=6.5cm] {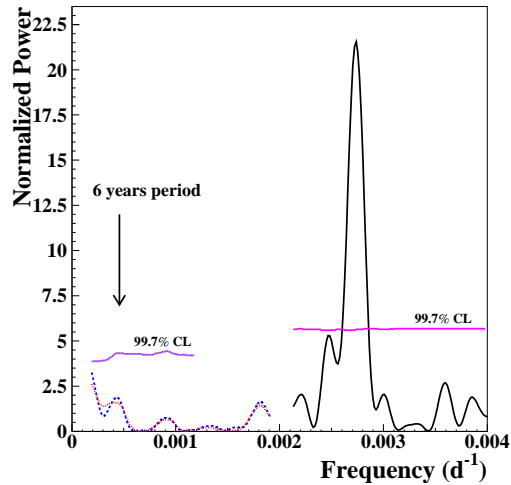}
\end{center}
\vspace{-0.6cm}
\caption{({\it Color online}).
Power spectrum of the annual baseline counting rates for (2--4) keV (dashed -- blue -- curve) and 
(2--6) keV (dotted -- red -- curve) {\it single-hit} events. Also shown for comparison is the power spectrum 
obtained by considering the 13 annual cycles of DAMA/NaI and DAMA/LIBRA for the {\it single-hit} 
residuals in (2--6) keV (solid -- black -- line) \cite{modlibra2}. The calculation has been performed according to 
ref. \cite{Lomb}, including also the treatment of the experimental errors and of the time binning. As can 
be seen, a principal mode is present at a frequency of $2.735 \times 10^{-3}$ d$^{-1}$, that corresponds to a 
period of $\simeq$ 1 year. The 99.7\% confidence lines for excluding the white noise hypothesis are also shown (see text). 
No statistically-significant peak is present at lower frequencies and, in particular,
at frequency corresponding to a 6-year period. This implies that no evidence for a long term modulation in the 
counting rate is present.}
\label{fg:lomb}
\end{figure}

\vspace{0.4cm}

For each annual cycle of DAMA/NaI and DAMA/LIBRA, we calculated 
the annual baseline counting rates -- that is the averages on all the detectors ($j$ index) of
$flat_{j}$ (i.e. the {\it single-hit} rate of the $j$-th detector averaged over the annual cycle,
see e.g. \cite{modlibra}) --
for the (2--4) keV and (2--6) keV energy intervals, respectively.
Their power spectra (dashed -- blue online -- and dotted -- red online -- curves, respectively) 
in the frequency range 0.0002--0.0018 d$^{-1}$ (corresponding to a period range
13.7--1.5 year) are reported in Fig. \ref{fg:lomb}; 
the power spectrum (solid black line) above 0.0022 d$^{-1}$, obtained when considering the 
(2--6) keV {\it single-hit} residuals of Fig. 1 of ref. \cite{modlibra2}, is reported
for comparison. 
To evaluate the statistical significance of these power spectra we have performed a Montecarlo
simulation imposing constant null expectations for residuals;
from the simulated power spectra the probability that an apparent periodic modulation may appear 
as a result of pure white noise has been evaluated.
The 99.7\% confidence lines for excluding the white noise hypothesis
are shown in Fig. \ref{fg:lomb}. 
A principal mode is present in the power spectrum of the experimental data
for a frequency equal to $2.735 \times 10^{-3}$ d$^{-1}$ (black solid curve), 
corresponding to a period of $\simeq$ 1 year, while no statistically-significant peak 
is present at lower frequencies and, in particular, at frequency 
corresponding to a 6-year period. 

\begin{figure}[!ht]
\begin{center}
\includegraphics[width=6.0cm] {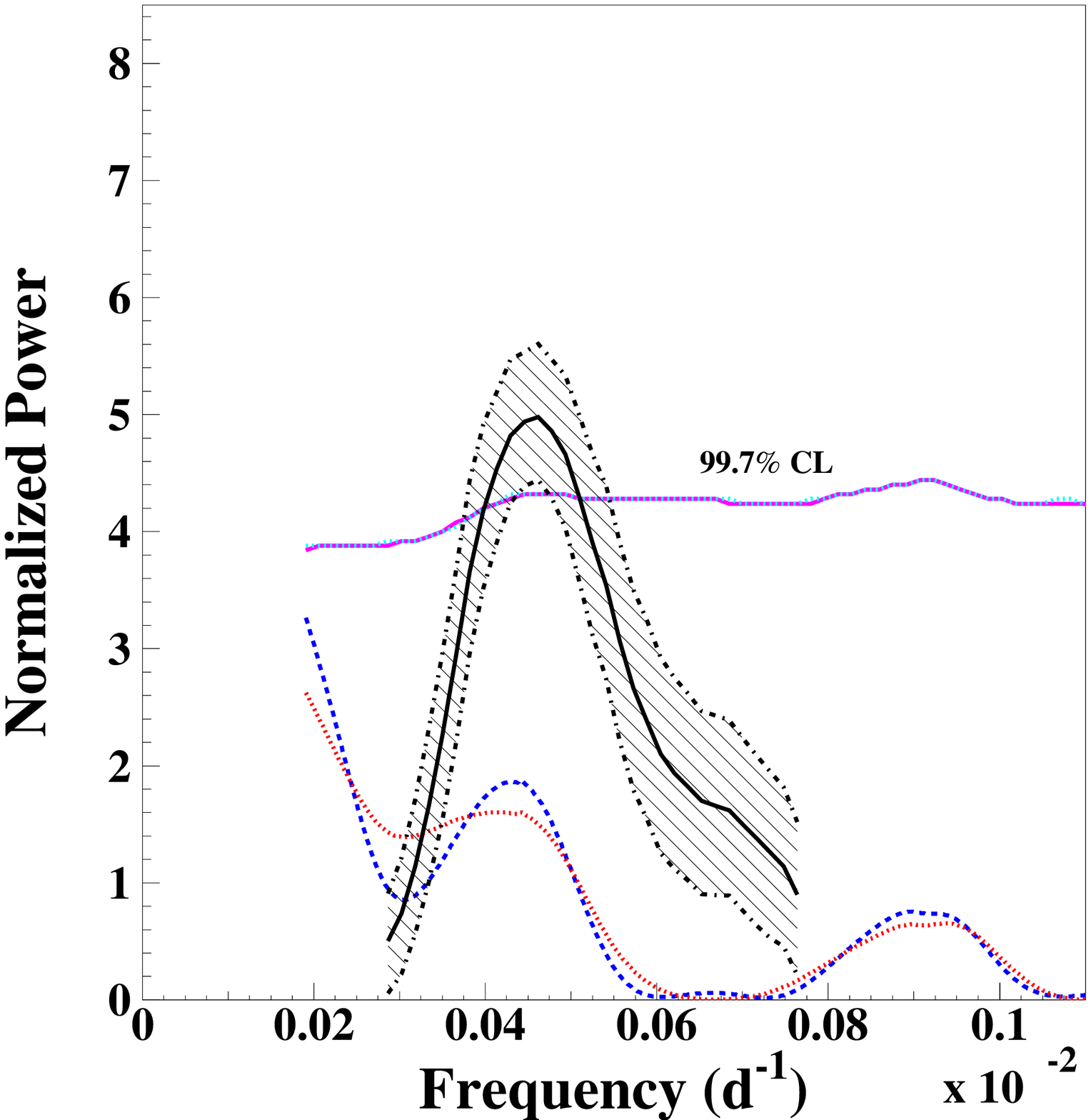}
\end{center}
\vspace{-0.6cm}
\caption{({\it Color online}). A detail of Fig. \ref{fg:lomb}, with superimposed power spectrum
(solid black curve),
expected in the hypothesis of a contribution from muons. 
The shaded region is the $1 \sigma$ (68\% C.L.) band.
The peak at 6 year period would be in this hypothesis well evident and above the threshold 
of detectability at $3\sigma$ C.L.. On the contrary, the power spectrum of the 
experimental data (dashed -- blue -- and dotted -- red -- curves)
is completely outside the $1 \sigma$ band.
For simplicity, the calculations are shown just for the cumulative (2--6) keV energy interval. 
This further shows that no evidence for a long term modulation in the 
counting rate is present, as it should already be expected on the basis of 
the many other arguments discussed in this paper. 
See text.}
\label{fg:lomb_cl}
\end{figure}

A further investigation of any hypothetical 6-year period has been performed, taking into account
that the LVD muon data have a 1-year period 
modulation amplitude equal to 1.5\% \cite{LVD}, and -- according to the claim of ref. \cite{blum} --
a 6-year period 
modulation amplitude equal to 1\% (actually
$\gsim 1 \%$, as reported in ref. \cite{blum}). 
Thus, in the case that muons might contribute to the DAMA effect, a 6-year period 
modulation would be present in the DAMA data with amplitude $\simeq 0.008$ cpd/kg/keV,
that is a $1\%/1.5\%$ fraction of the 1-year period modulation amplitude measured by DAMA.
A simulation of $10^6$ experiments 
has been performed and the power spectrum of each simulation has been derived. 
The average of all the simulated power spectra is reported in Fig. \ref{fg:lomb_cl}
as solid black curve; the shaded region is the $1 \sigma$ (68\% C.L.) band.
The hypothetical peak at 6-year period would be under these assumptions well evident and above the threshold 
of detectability at $3\sigma$ C.L.. On the contrary, the power spectrum of the 
experimental data is completely outside the $1 \sigma$ band.
For simplicity, these calculations have been reported just for the cumulative (2--6) keV energy interval. 

This further shows that no evidence for a long term modulation in the counting rate is 
present, as -- on the other hand -- it should already be expected on the basis of 
the many other arguments (and just one suffices) discussed in this paper,
further demonstrating that there is no role for muons.

\subsection{No role for muons from statistical considerations}

In addition to the previous arguments, let us
finally demonstrate that any hypothetical effect -- even with high-multiplicity production -- due to muons crossing 
the NaI(Tl) 
detectors and/or the surroundings of the set-up cannot give any appreciable contribution to the observed (2--6) keV 
{\it single-hit} event rate, just owing to statistical considerations. In fact, because of the poissonian 
fluctuation on the number of muons, the standard deviation, $\sigma(A)$, of any hypothetically induced (2--6) keV 
{\it single-hit} modulation amplitude, $A$, would be much larger than measured by DAMA, thus, giving rise to no 
statistically-significant effect.
To explain this argument, Fig. \ref{fg:stat} (see Appendix) reports the expected $\sigma(A)$ (solid line) as a function 
of an effective area, $A_{eff}$, around the set-up.
This effective area is defined as the area crossed by the muons that would give rise 
to the bulk contribution (say 
90\% of the total) of the DAMA measured annual modulation amplitude through their ``products'' 
(gamma's, beta's, neutrons, phosphorescence pulses, whatever exotics, ...). Therefore, muons outside this area can 
safely be neglected for the purpose of this description. As a matter of fact, the size of the effective area 
depends on the considered ``product'' of muons, on its nature and on its interactions with the materials around 
the DAMA/LIBRA set-up. The smaller is $A_{eff}$, the smaller would be the number of muons that might 
hypothetically contribute (either directly or indirectly) to the DAMA annual modulation amplitude, and the larger 
would be the fluctuation $\sigma(A)$. 

\begin{figure}[!ht]
\begin{center}
\includegraphics[width=8.cm] {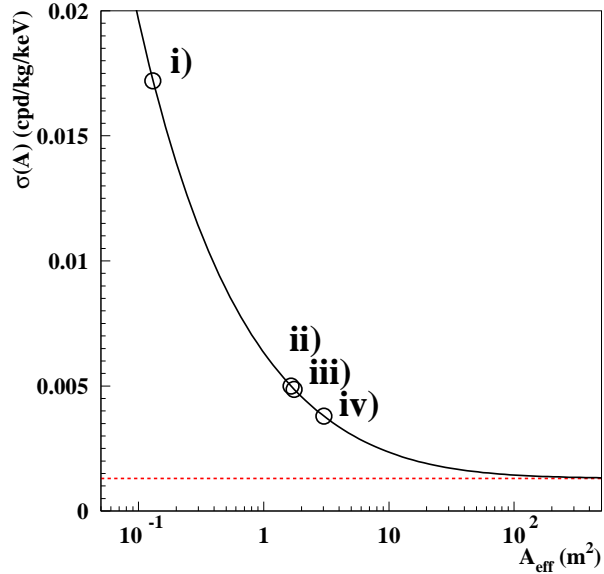}
\end{center}
\vspace{-0.8cm}
\caption{{\it (Color online)}.
Expected standard deviation (solid line), $\sigma(A)$, of the (2--6) keV {\it single-hit} annual modulation amplitude, $A$, as 
a function of the effective area, $A_{eff}$, in the hypothetical case that muons might produce in $A_{eff}$ 
``something'' (gamma's, beta's, neutrons, phosphorescence pulses, exotics, ...) able to contribute to the DAMA (2--6) 
keV {\it single-hit} modulation amplitude (see text and Appendix). The value experimentally observed by DAMA is 
shown as (red) dashed line; the $\sigma(A)$ curve approaches this value just for $A_{eff} \gg 50$ m$^2$ while even in 
the most cautious case the $A_{eff}$ value is much smaller. The four interesting cases described in the text are shown; 
as it can be seen, in all four cases the fluctuation is much larger than that observed by DAMA.}
\label{fg:stat}
\end{figure}

The value experimentally observed by DAMA for $\sigma(A)$, 0.0013 cpd/kg/keV \cite{modlibra2}, is shown as (red online) 
dashed line in Fig. \ref{fg:stat}; the $\sigma(A)$ curve approaches this value just for $A_{eff} \gg 50$ m$^2$, while 
-- as we will demonstrate -- even in the most cautious case the $A_{eff}$ value is much smaller. In fact, to have an idea 
on the possible sizes of $A_{eff}$, four interesting cases are shown in Fig. \ref{fg:stat}:
i) the muons interacting directly in the NaI(Tl) DAMA/LIBRA detectors (hypothetically producing either very short range 
particles or phosphorescence pulses as discussed in section \ref{phos}, ...), corresponding to $A_{eff}$ equal to the 
DAMA/LIBRA exposed surface: 0.13 m$^2$;
ii) the effective area equal to the one calculated (by a Montecarlo simulation) just considering
the $1/r^2$ dependence for the flux of the ``products'', without including 
any shield effect, corresponding to $A_{eff} \simeq 1.65$ m$^2$;
iii) the effective area equal to the area of the heavy passive shield, $A_{eff} \simeq 1.75$ m$^2$; 
iv) the effective area equal to the area of the heavy passive shield plus the neutron moderator, $A_{eff} \simeq 3$ m$^2$.
In all the four cases the fluctuation, driven by the small number of muons, is much larger than that observed by DAMA.
Since it is extremely safe to consider that any hypothetical mechanism would have a corresponding $A_{eff}$ within the 
previous considered cases (that is $A_{eff} \ll 50$ m$^2$), we can conclude that all (standard and exotic) mechanisms, 
because of the low number of the involved muons, provide too high fluctuations of the data, not observed in DAMA.
Even just this argument is enough to discard any kind of hypothesis about muons.

\section{Conclusion}

In this paper we have compiled some of the main scientific and quantitative arguments which demonstrate that there is no 
room for any hypothetical contribution from muons to the (2--6) keV {\it single-hit} annual modulation amplitude measured 
by DAMA experiments.
Some comments about incorrect arguments reported in recent papers 
\cite{ralston,nygren,blum,hongbo,cline} have also been addressed.
In conclusion, the hypothesis of a role for muons in the DAMA observed (2--6) keV 
{\it single-hit} annual modulation can be safely rejected for many scientific reasons.

\section{Appendix}

We give here some elements to properly evaluate the expected standard deviation
of the annual modulation amplitude in an experiment studying the DM annual modulation
signature, as DAMA does. Two results are compared
with the experimental values and the curve of Fig. \ref{fg:stat} is justified. 

Let us simplify the result of an annual modulation experiment as a set of
$N_i$ counts per each day $i$ in the interesting energy region. The index $i$ identify the day 
in the year and, therefore, ranges from 0 to 365.
The experimental information about the data taking can be gathered in 
another quantity that is the set of the daily exposure $w_i=M \Delta t_i \Delta E \eta$, 
where $M$ is the exposed mass, $\Delta t_i$ is the live time during the day, $t_i$ is the 
time of the bin (center-bin value),
$\Delta E$ is the energy bin and $\eta$ is the efficiency. A new variable can be defined:
\begin{equation}
 X = \Sigma_i N_i (c_i-\beta),
\end{equation}
where: i) $c_i = cos \omega (t_i-t_0)$; ii) $\omega=\frac{2 \pi}{T}$; iii) T is the period (1 year);
iv) $t_0$ is
the phase of the annual modulation ($t_0 = 152.5$ day of the year). 
Moreover, it turns out useful to define the following quantities:
the total exposure $W=\Sigma_i w_i$, the mean value of the cosine function 
$\beta = \frac{\Sigma_i w_i c_i}{W}$, and the variance of the cosine function 
$(\alpha - \beta^2) = \frac{\Sigma_i w_i (c_i - \beta)^2}{W}$. For a data taking
involving the whole year, one can expect $\beta \simeq 0$ and $(\alpha - \beta^2) \simeq 0.5$.

The variable $X$ allows us to obtain information about the parameters of the
annual modulation. In fact, from the experimental data one can obtain the value of $X$ 
and its error $\sigma_X = \sqrt{Var(X)}$; the variance can be calculated directly from the
experimental data hypothesizing that $N_i$ is distributed as poissonian variable
(that is $Var(N_i)=N_i$):
\begin{equation}
Var(X) = \Sigma_i N_i (c_i-\beta)^2. 
\label{eq:varx}
\end{equation}
These two quantities
are directly connected with the parameters of the annual modulation; see later.

In the following two cases will be considered: i) the case of an 
annual modulation induced by DM particles; ii) the case of an
annual modulation due to ``generic products'' of the muons
crossing the set-up and/or the surroundings.

\vspace{0.4cm}
{\tt \bf Case of annual modulation induced by DM particles.}
In such a case the expectation of the stochastic variable $N_i$ is given by:
\begin{equation}
E(N_i) = (b + S_0 + S_m c_i) \times w_i. 
\end{equation}
Here $b$ is the background rate,
$S_0$ is the unmodulated component of the DM signal and $S_m$ is the modulation amplitude.
The expected value of the $X$ variable is connected with the modulation amplitude by
\begin{equation}
E(X) = \Sigma_i E(N_i) \cdot (c_i-\beta) = 
(b+S_0) \cdot \Sigma_i w_i (c_i-\beta) + S_m \cdot \Sigma_i w_i c_i (c_i-\beta),
\end{equation}
that is, since the first term is null because of the definition of $\beta$:
\begin{equation}
E(X) = S_m W (\alpha - \beta^2).
\end{equation}
Therefore, an estimate of $X$ through the experimental data gives the $S_m$ evaluation.
The sensitivity can be determined by considering that 
$Var(N_i) \simeq (b + S_0)w_i$, since the number of events have a
poissonian distribution and the $ S_m c_i$ terms can be safely neglected. 
Hence, the variance of $X$ can be written from eq. (\ref{eq:varx}) as:
\begin{equation}
Var(X) \simeq \Sigma_i (b + S_0)w_i (c_i-\beta)^2 =
(b + S_0) W (\alpha - \beta^2).
\end{equation}
The sensitivity reachable on the modulation amplitude is given by:
\begin{equation}
\sigma(S_m) = \frac{\sqrt{Var(X)}}{W(\alpha-\beta^2)} \simeq \sqrt{\sigma_B^2 + \sigma_{DM}^2},
\end{equation}
with $\sigma_B^2 = \frac{b}{W(\alpha-\beta^2)}$ and $\sigma_{DM}^2 = \frac{S_0}{W(\alpha-\beta^2)}$.

Let us now consider the case of the DAMA/NaI (target mass $\simeq 100$ kg)
and DAMA/LIBRA (target mass $\simeq 250$ kg) results: 
$M \cdot T = 425428 $ kg day, $\Delta E = 4 $ keV, 
$\eta \simeq 0.7$, $W (\alpha-\beta^2) = 5.96 \times 10^5$ kg day keV.
The counting rate of the set-ups ($b+S_0$) is around 1 cpd/kg/keV.
Thus, $\sigma(S_m) \simeq 0.0013$ cpd/kg/keV, well comparable with the one obtained by DAMA.
Therefore, such a sensitivity has allowed the measurement of a modulation amplitude $S_m$ of the order of 
$10^{-2}$ cpd/kg/keV.
We remind that the {\it single-hit} modulation amplitude in the (2--6) keV energy region measured by DAMA/NaI and DAMA/LIBRA is 
$(0.0114 \pm 0.0013)$ cpd/kg/keV (here period and phase fixed in the fit).

In Fig. \ref{fg:stat} the experimental value of $\sigma(S_m)$ and the expected $\sigma(S_m)$ are 
reported as a (red online) dashed line; actually, they are overlapped. 

In conclusion, the experimental value of $\sigma(S_m)$ is well compatible
with the expectation, giving further support to the evidence for an 
annual modulation effect induced by DM particles.

\vspace{0.4cm}
{\tt \bf Case of annual modulation directly or indirectly due to muons
crossing the set-up and/or the surroundings.}
We consider here the case of hypothetical effects 
correlated with muons crossing the NaI(Tl) detectors 
and/or the surroundings of the set-up.

To easily describe the model, an effective area, $A_{eff}$, can be defined as 
the area around the DAMA/LIBRA set-up
where whatever (even hypothetical) product of muons (gamma's, beta's, neutrons, phosphorescence pulses, 
any exotics, ...) would give the bulk contribution (we would say 90\% of the total) 
to the measured (2--6) keV {\it single-hit} modulation; 
muons outside this area are for simplicity not considered in the following,
because of their slight contribution.

The rate of muons inside the effective area is $r_\mu = \Phi_{\mu} \times A_{eff}$ muons/day and 
its relative annual modulation is $\Delta \simeq 0.015$ \cite{LVD,borexino}.
Let us assume for a while as true the scenario: i)
where each muon, crossing the NaI(Tl) detectors and the $A_{eff}$ area, might produce during 
the incoming period (minutes, hours, days)
$\varepsilon$ {\it single-hit} events in the considered low energy bin in all the DAMA detectors;
ii) the period and the phase of the muon flux is equal to those of the DM signal. 
By the way, $\varepsilon$ is distributed as
a poissonian variable with expectation and variance equal to $\bar{\varepsilon}$.

The expected value of $N_i$ would be in this case (for simplicity $\eta \sim 1$ is assumed):
\begin{equation}
E(N_i) = (b + R_\mu \cdot \bar{\varepsilon} 
+ \Delta \cdot R_\mu \cdot \bar{\varepsilon} \cdot c_i) \times w_i; 
\end{equation}
here $b$ is still the background and $R_\mu = \frac{r_\mu}{M \Delta E}$. 
Now considering the results obtained by DAMA and the assumed scenario, an estimation
of the $\bar{\varepsilon}$ parameter can be obtained from the measured modulation amplitude, 
$A \simeq 10^{-2}$ cpd/kg/keV: 
$\bar{\varepsilon} = \frac{A}{\Delta \cdot R_\mu} \simeq \frac{33}{A_{eff}[\textrm{m}^2]}$.
For example, if $A_{eff} = 0.13$ m$^2$, as for the case of direct interaction of muons
producing e.g. hypothetical delayed phosphorescence pulses \cite{nygren} (hypothesis already
discarded above), 
each muon should have to give rise to $\bar{\varepsilon} \simeq 255$ 
{single-hit} (2--6) keV events in all the DAMA/LIBRA set-up,
value in agreement with that given in Sect. \ref{phos}.
Moreover, $ R_\mu \cdot \bar{\varepsilon} \simeq \frac{A}{\Delta}\simeq 0.68$ cpd/kg/keV;
considering that the total counting rate of 
the DAMA detectors is around $\sim 1$ cpd/kg/keV, one can derive $b \simeq 0.32$ cpd/kg/keV. 

Following the same procedure as above, one gets:
\begin{equation}
E(X) = \Delta \cdot R_\mu \cdot \bar{\varepsilon} \cdot W (\alpha - \beta^2),
\end{equation}
and
\begin{equation}
Var(X) = \Sigma_i (bw_i + R_\mu w_i \bar{\varepsilon}^2 + R_\mu w_i \bar{\varepsilon}) (c_i-\beta)^2.
\label{eq:varx2}
\end{equation}
To obtain the latter equation one can profit of the fact that 
$N_i$ can be written as the sum of two components, the number of background events and the number of the
events hypothetically induced by muons:
$N_i= N_i^B + N_i^\mu$, with $E(N_i^B)=Var(N_i^B)=b w_i$. 

The stochastic variable $N_i^\mu$
can be written as $N_i^\mu = \displaystyle\sum\limits_{k=1}^m \mathcal{E}_k$, where $m$ 
is the number of muons crossing the $A_{eff}$ area and the $\mathcal{E}_k$ are the numbers of low-energy events
produced by the $k$-th muon. Of course, $E(m)=Var(m)=R_\mu w_i$ and 
$E(\mathcal{E}_k)=Var(\mathcal{E}_k)=\bar{\varepsilon}$. 
Thus, one can write:
\begin{equation}
E(N_i^\mu)=  \displaystyle\sum\limits_{k=1}^{E(m)} E(\mathcal{E}_k) = R_\mu w_i \bar{\varepsilon}.
\end{equation}
and the $Var(N_i^\mu)$ can be calculated, considering that $N_i^\mu$ 
has two sources of fluctuation: i) one associated to the number of crossing muons;
ii) one associated to the hypothetical production of low-energy events.
Without loosing generality, the two terms can be written as:
\begin{equation}
Var(N_i^\mu) \simeq Var(m) \cdot \left(\frac{\partial}{\partial m} \displaystyle\sum\limits_{k=1}^m 
E(\mathcal{E}_k)
\right)^2 + \displaystyle\sum\limits_{k=1}^{E(m)} Var(\mathcal{E}_k) = R_\mu w_i \bar{\varepsilon}^2 +
R_\mu w_i \bar{\varepsilon}.
\end{equation}
All this justifies eq. (\ref{eq:varx2}). One can see that for large number of $\bar{\varepsilon}$
the contribution of the fluctuation of the number of muons is largely dominant.

Finally, the sensitivity reachable on the modulation amplitude, $A$, is given by:
\begin{equation}
\sigma(A) = \frac{\sqrt{Var(X)}}{W(\alpha-\beta^2)} \simeq \sqrt{\sigma_B^2 + \sigma_\mu^2 + \sigma_\varepsilon^2},
\label{eq:sigA}
\end{equation}
where $\sigma_B^2=b/\left[W (\alpha-\beta^2)\right]$ takes into account the fluctuation of the number of the background events,
$\sigma_\mu^2=\left(R_\mu \bar{\varepsilon}^2\right)/\left[W (\alpha-\beta^2)\right]$ 
the fluctuation of the number of muons crossing the $A_{eff}$ area and, finally,
$\sigma_\varepsilon^2=\left(R_\mu \bar{\varepsilon} \right)/\left[W (\alpha-\beta^2)\right]$ 
takes into account the fluctuation of the number of the hypothetically produced low energy events after a muon.

Considering for the parameters the values given above, one has:
\begin{equation}
\sigma_B            = 7.3 \times 10^{-4} \textrm{ cpd/kg/keV};
\end{equation}
\begin{equation}
\sigma_\mu          = \frac{6.2 \times 10^{-3}}{\sqrt{A_{eff}(\textrm{m}^2)}} \textrm{ cpd/kg/keV};
\end{equation}
\begin{equation}
\sigma_\varepsilon  = 1.07 \times 10^{-3} \textrm{ cpd/kg/keV}.
\end{equation}

Therefore, the major contribution to the error for $A_{eff} < 100$  m$^2$ 
is from the fluctuation of the number of muons crossing the NaI(Tl) detectors and the $A_{eff}$ area;
in addition, this contribution becomes larger for smaller $A_{eff}$, since the smaller would be
the number of muons hypothetically able to directly or indirectly contribute to the DAMA (2--6) keV {\it single-hit} 
modulation amplitude,
the larger would be the fluctuation $\sigma(A)$. 
The standard deviation $\sigma(A)$, calculated as in eq. (\ref{eq:sigA}), 
is reported in Fig. \ref{fg:stat} as function of the effective area $A_{eff}$.

For example, if $A_{eff}$ is the exposed NaI(Tl) surface of DAMA/LIBRA, the standard deviation 
$\sigma(A)$ is in this case equal to 0.017 cpd/kg/keV, that is 13 times larger
than the one measured by DAMA.

Other enlightening cases, reported in the text and in Fig. \ref{fg:stat}, show effective areas at level of few m$^2$;
in all these cases the fluctuations are much larger than the observed value by DAMA.
Since it is extremely safe to assume that any hypothetical mechanism would have 
a corresponding $A_{eff}$ within the previous considered cases, we can 
conclude that all (standard and exotic) mechanisms, because of the low number of the involved
muons, provide too high fluctuations of the data, not observed in DAMA.

In conclusion any hypothetical and quantitatively appreciable 
effect correlated with muons crossing either the NaI(Tl) detectors 
or the surroundings of the set-up can be further excluded, 
even already just owing to statistical considerations about 
the poissonian fluctuation on the number of muons.

In other words, because of the low number of the involved muons, 
all (standard and exotic) muon-induced mechanisms
provide fluctuations of the data larger than those observed by DAMA, and therefore
-- in addition to all the other arguments given in this paper -- 
cannot give rise to any evidence of modulation in DAMA.

\end{document}